\documentclass[pdftex,twocolumn,epjc3,a4paper]{svjour3}

\RequirePackage[T1]{fontenc}
\smartqed  
\usepackage{amsmath}
\usepackage{amssymb}
\RequirePackage{mathptmx}      
\RequirePackage{flushend}
\RequirePackage{mathrsfs}
\RequirePackage{dsfont}
\usepackage{units}
\usepackage{graphicx}
\usepackage{cite}
\usepackage{xcolor} 
\RequirePackage[colorlinks,citecolor=blue,urlcolor=blue,linkcolor=blue]{hyperref}

\newcommand{\GeV}{GeV/$c^2$}

\journalname{Eur. Phys. J. C}

\begin{document}

\author
{%
G.~Angloher\thanksref{addr1}\and
P.~Bauer\thanksref{addr1}\and
A.~Bento\thanksref{addr1,addr2}\and
E.~Bertoldo\thanksref{addr1} \and
C.~Bucci\thanksref{addr3}\and
L.~Canonica\thanksref{addr1}\and
A.~D'Addabbo\thanksref{addr3,addr10}\and
X.~Defay\thanksref{addr4}\and
S.~Di Lorenzo\thanksref{addr3,addr10}\and
A.~Erb\thanksref{addr4,addr5}\and
F.~v.~Feilitzsch\thanksref{addr4}\and
N.~Ferreiro~Iachellini\thanksref{addr1}\and
P.~Gorla\thanksref{addr3}\and
D.~Hauff\thanksref{addr1}\and
J.~Jochum\thanksref{addr6}\and
M.~Kiefer\thanksref{addr1}\and
H.~Kluck\thanksref{addr7,addr8}\and
H.~Kraus\thanksref{addr9}\and
A.~Langenk\"amper\thanksref{addr4}\and
M.~Mancuso\thanksref{addr1}\and
V.~Mokina\thanksref{addr7}\and
E.~Mondragon\thanksref{addr4}\and
V.~Morgalyuk\thanksref{addr4}\and
A.~M\"unster\thanksref{addr4}\and
M.~Olmi\thanksref{addr3,addr10}\and
C.~Pagliarone\thanksref{addr3,addr11}\and
F.~Petricca\thanksref{addr1}\and
W.~Potzel\thanksref{addr4}\and
F.~Pr\"obst\thanksref{addr1}\and
F.~Reindl\thanksref{addr7,addr8}\and
J.~Rothe\thanksref{addr1}\and
K.~Sch\"affner\thanksref{addr3,addr10}\and
J.~Schieck\thanksref{addr7,addr8}\and
V.~Schipperges\thanksref{t1,e1,addr6}\and
S.~Sch\"onert\thanksref{addr4}\and
M.~Stahlberg\thanksref{addr7,addr8}\and
L.~Stodolsky\thanksref{addr1}\and
C.~Strandhagen\thanksref{addr6}\and
R.~Strauss\thanksref{addr1}\and
C.~T\"urkoglu\thanksref{addr7,addr8}\and
I.~Usherov\thanksref{addr6}\and
M.~Willers\thanksref{addr4}\and
M.~W\"ustrich\thanksref{addr1}\and
V.~Zema\thanksref{addr3,chalmers,addr10}\\
(The CRESST Collaboration) \\ and \\
R.~Catena\thanksref{t1,e2,chalmers}
}
\institute
{%
Max-Planck-Institut f\"ur Physik, D-80805 M\"unchen, Germany \label{addr1} \and
INFN, Laboratori Nazionali del Gran Sasso, I-67010 Assergi, Italy \label{addr3} \and
Physik-Department and Excellence Cluster Universe, Technische Universit\"at M\"unchen, D-85748 Garching, Germany \label{addr4} \and
Eberhard-Karls-Universit\"at T\"ubingen, D-72076 T\"ubingen, Germany \label{addr6} \and
Institut f\"ur Hochenergiephysik der \"Osterreichischen Akademie der Wissenschaften, A-1050 Wien, Austria \label{addr7} \and
Atominstitut, Vienna University of Technology, A-1020 Wien, Austria \label{addr8} \and
Department of Physics, University of Oxford, Oxford OX1 3RH, United Kingdom \label{addr9} \and
Chalmers University of Technology, Department of Physics, SE-412 96 G\"oteborg, Sweden \label{chalmers} \and
also at: GSSI-Gran Sasso Science Institute, 67100, L'Aquila, Italy \label{addr10} \and
also at: Departamento de Fisica, Universidade de Coimbra, P3004 516 Coimbra, Portugal \label{addr2} \and
also at: Walther-Mei\ss ner-Institut f\"ur Tieftemperaturforschung, D-85748 Garching, Germany \label{addr5} \and
also at: Dipartimento di Ingegneria Civile e Meccanica, Universit\'{a} degli Studi di Cassino e del Lazio Meridionale, I-03043 Cassino, Italy \label{addr11}
}
\thankstext[$\star$]{t1}{Corresponding author}
\thankstext{e1}{vincent.schipperges@uni-tuebingen.de}
\thankstext{e2}{catena@chalmers.se}

\title{Limits on Dark Matter Effective Field Theory Parameters with CRESST-II}

\date{\today}

\maketitle
\begin{abstract}
CRESST is a direct dark matter search experiment, aiming for an observation of nuclear recoils induced by the interaction of dark matter particles with cryogenic scintillating calcium tungstate crystals. Instead of confining ourselves to standard spin-independent and spin-dependent searches, we re-analyze data from CRESST-II using a more general effective field theory (EFT) framework. On many of the EFT coupling constants, improved exclusion limits in the low-mass region (< \unit[3-4]{\GeV}) are presented.
\end{abstract}

\section{Introduction}

The elusive nature of dark matter remains one of the major unsolved mysteries in modern physics. One leading hypothesis is that dark matter consists of as yet undetected particles with interactions at the weak scale (or below)~\cite{Bertone:2004pz}. If the hypothesis is correct, the microscopic properties of dark matter might be revealed in the coming years using existing detection methods~\cite{Bertone:2010at}. The direct detection technique will play a key role in this context~\cite{Baudis:2012ig}. It searches for nuclear recoils induced by the non-relativistic scattering of Milky Way dark matter particles in low-background detectors~\cite{Drukier:1983gj,Goodman:1984dc}. Detectors based on dual-phase time projection chambers have proven to be very effective in the search for dark matter particles heavier than about \unit[10]{\GeV}~\cite{Undagoitia:2015gya}. WIMPs (for Weakly Interacting Massive Particles) are the leading dark matter candidate in this mass range~\cite{Arcadi:2017kky}. On the other hand, when the dark matter particle mass is below few \GeV, cryogenic experiments provide the best sensitivity to dark matter-nucleon interactions because of the low energy threshold these detectors can achieve~\cite{Undagoitia:2015gya}. The experiment CRESST (for Cryogenic Rare Event Search with Superconducting Thermometers) has pioneered the search for sub-\GeV~dark matter, and currently places the most stringent exclusion limits on the spin-independent dark matter-nucleon scattering cross-section for dark matter masses below \unit[1.8]{\GeV}~\cite{Petricca:2017zdp}. Dark matter in the \GeV~mass range is expected in models where the present cosmological density of dark matter is explained in terms of freeze-in and asymmetric production (for a review, see~\cite{Battaglieri:2017aum}).

From the theoretical side, the null result of present direct detection experiments is usually interpreted within a framework where dark matter either couples to the total nucleon content of the nucleus (spin-independent interaction) or to the nucleon spin content of the nucleus (spin-dependent interaction)~\cite{Lewin:1995rx}. This is a reasonable approach, since spin-independent and spin-dependent interactions are in general expected to give the leading contribution to the cross section for dark matter-nucleus scattering. However, when these standard interactions are forbidden or suppressed, such as in the case of dark matter-nucleon interactions mediated by a pseudo-scalar particle~\cite{Arina:2014yna} or in the case of anapole interactions~\cite{DelNobile:2014eta}, the leading contribution to the dark matter-nucleus scattering cross section may have a different nature. The classification and characterisation of non-standard dark matter-nucleus interactions has driven the theoretical research in the field of dark matter direct detection in the past few years. In this context, the non-relativistic effective theory of dark matter-nucleon interactions has played a key role~\cite{Fitzpatrick:2012ix}. Assuming that dark matter and nucleons are the only relevant degrees of freedom, this theory describes all possible dark matter-nucleon interactions which are compatible with the symmetries characterising the non-relativistic dark matter-nucleus scattering. Within this theoretical framework, data collected by the SuperCDMS and XENON100 experiments have been interpreted in~\cite{Schneck:2015eqa} and \cite{Aprile:2017aas}, respectively. Furthermore, a likelihood analysis of different direct detection experiments and non-relativistic dark matter-nucleon interactions has been performed in~\cite{DelNobile:2013sia,Catena:2014uqa,Rogers:2016jrx,Kang:2018odb}. The role of operator interference has extensively been discussed in~\cite{Catena:2015uua}.

The main goal of this work is to set exclusion limits on the coupling constants of the effective theory of dark matter-nucleon interactions using data collected by the CRESST-II Phase 2 experiment. This analysis generalises previous results found by the CRESST collaboration focusing on standard interactions, and extends limits on non-standard interactions presented by other groups to an as yet unexplored mass range. 

This article is organised as follows. We introduce the non-relativistic effective theory of dark matter-nucleon interactions in Sec.~\ref{eftdm} and the CRESST experiment in Sec.~\ref{cresst}. Sec.~\ref{analysis} is devoted to the methods and data used in our analysis, while a summary of our results and conclusions is presented in Sec.~\ref{conclusions}.
\newpage

\section{Effective Theory of Dark Matter Direct Detection}\label{eftdm}
In this section we briefly review the non-relativistic effective theory of dark matter-nucleon interactions as defined in ~\cite{Fitzpatrick:2012ix}.~The theory is based upon the following considerations:~1)~In the non-relativistic limit, the amplitude for dark matter scattering off nucleons $N$ in target nuclei, $\mathscr{M}_{\chi N}$, can in general be expanded in powers of $|\vec q|/m_{N}\ll 1$, where $|\vec q|$ is the momentum transferred in the scattering and $m_N$ is the nucleon mass.~2)~Each term in this expansion must be invariant under Galilean transformations and Hermitian conjugation, and can be expressed in terms of basic invariants under the above symmetries~\cite{Fitzpatrick:2012ix}:~$\vec S_\chi$, $\vec S_N$, $i\vec q$, and $\vec v^\perp \equiv \vec v + \vec q/2 \mu_N$, where $\vec S_{\chi}$ ($\vec S_N$) is the dark matter (nucleon) spin, and $\mu_N$ and $\vec v$ are the dark matter-nucleon reduced mass and relative velocity, respectively.~These considerations imply that in the one-body approximation\footnote{The one-body approximation assumes that the interactions of dark matter with nucleon pairs can be neglected, i.e.~the so-called two-body currents.~See~\cite{Hoferichter:2015ipa} for a discussion on two-body currents in dark matter-nucleus scattering.}, 
the Hamiltonian for the interactions of dark matter with a nucleus $T$, $\mathscr{H}_{\chi T}$, can be written as follows~\cite{Fitzpatrick:2012ix}:
\begin{equation}
\mathscr{H}_{\chi T} = \sum_{i}\sum_{j} \Big( c_j^0  \hat{\mathcal{O}}^i_j \, \mathds{1}_{2\times 2}^i + c_j^1 \, \hat{\mathcal{O}}^i_j \, \mathds{\tau}_{3}^i \Big) \,,
\label{eq:H}
\end{equation}
where the index $j$ characterises the dark matter-nucleon interaction type and $c_j^0$ ($c_j^1$) is the associated isoscalar (isovector) coupling constant.~The $A$ nucleons in the target nucleus are labeled by the index $i=1,\dots,A$, and $\mathds{1}_{2\times 2}^i$ ($\mathds{\tau}_{3}^i$) is the identity (third Pauli matrix) in the $i$-th nucleon isospin space. Finally, $\hat{\mathcal{O}}^i_j$ is a non-relativistic operator for interactions of type $j$ between the dark matter particle and the $i$-th nucleon in the nucleus. 

In the Hamiltonian (\ref{eq:H}), the operators $\hat{\mathcal{O}}^i_j$ act on particle coordinates through the momentum transfer and transverse relative velocity operators, $\hat{\vec q}$ and $\hat{\vec v}^\perp$, respectively.~At linear order in $\hat{\vec v}^{\perp}$, and at second order in $\hat{\vec q}$, Eq.~(\ref{eq:H}) includes 16 independent interaction operators $\hat{\mathcal{O}}^i_j$, listed in Tab.~\ref{tab:operators}. Not all of them appear as leading operators in the non-relativistic limit of simplified models\footnote{By ``simplified model'', one usually refers to a model where dark matter interacts with nucleons through the exchange of a single mediator particle.}~\cite{Dent:2015zpa,Baum:2017kfa,Brod:2017bsw}.~In Tab.~\ref{tab:operators}, the dark matter and nucleon spin operators are denoted by $\hat{\vec S}_\chi$ and $\hat{\vec S}_N$, respectively.~Following~\cite{Fitzpatrick:2012ix}, here we do not consider the operators $\hat{\mathcal{O}}^i_{2}$ and $\hat{\mathcal{O}}_{16}^i$.~Indeed, $\hat{\mathcal{O}}_{2}^i$ is quadratic in $\vec v^\perp$, and $\hat{\mathcal{O}}_{16}^i$ is a linear combination of $\hat{\mathcal{O}}_{12}^i$ and $\hat{\mathcal{O}}_{15}^i$.~Notably, $\hat{\mathcal{O}}_{17}^i$ and $\hat{\mathcal{O}}_{18}^i$ can only arise for spin 1 dark matter~\cite{Dent:2015zpa}.~For simplicity, from here onwards we will omit the nucleon index $i$ in the definitions.

\begin{table}[t]
    \centering
    \begin{tabular*}{\columnwidth}{@{\extracolsep{\fill}}llll@{}}
    \hline
        $\hat{\mathcal{O}}_1 = \mathds{1}_{\chi}\mathds{1}_N$  \\
        $\hat{\mathcal{O}}_3 = i{\bf{\hat{S}}}_N\cdot\left(\frac{{\bf{\hat{q}}}}{m_N}\times{\bf{\hat{v}}}^{\perp}\right)\mathds{1}_\chi$ \\
        $\hat{\mathcal{O}}_4 = {\bf{\hat{S}}}_{\chi}\cdot {\bf{\hat{S}}}_{N}$ \\
        $\hat{\mathcal{O}}_5 = i{\bf{\hat{S}}}_\chi\cdot\left(\frac{{\bf{\hat{q}}}}{m_N}\times{\bf{\hat{v}}}^{\perp}\right)\mathds{1}_N$ \\
        $\hat{\mathcal{O}}_6 = \left({\bf{\hat{S}}}_\chi\cdot\frac{{\bf{\hat{q}}}}{m_N}\right) \left({\bf{\hat{S}}}_N\cdot\frac{\hat{{\bf{q}}}}{m_N}\right)$\\
        $\hat{\mathcal{O}}_7 = {\bf{\hat{S}}}_{N}\cdot {\bf{\hat{v}}}^{\perp}\mathds{1}_\chi$ \\
        $\hat{\mathcal{O}}_8 = {\bf{\hat{S}}}_{\chi}\cdot {\bf{\hat{v}}}^{\perp}\mathds{1}_N$  \\
        $\hat{\mathcal{O}}_9 = i{\bf{\hat{S}}}_\chi\cdot\left({\bf{\hat{S}}}_N\times\frac{{\bf{\hat{q}}}}{m_N}\right)$ \\
        $\hat{\mathcal{O}}_{10} = i{\bf{\hat{S}}}_N\cdot\frac{{\bf{\hat{q}}}}{m_N}\mathds{1}_\chi$   \\
        $\hat{\mathcal{O}}_{11} = i{\bf{\hat{S}}}_\chi\cdot\frac{{\bf{\hat{q}}}}{m_N}\mathds{1}_N$   \\
        $\hat{\mathcal{O}}_{12} = {\bf{\hat{S}}}_{\chi}\cdot \left({\bf{\hat{S}}}_{N} \times{\bf{\hat{v}}}^{\perp} \right)$  \\
        $\hat{\mathcal{O}}_{13} =i \left({\bf{\hat{S}}}_{\chi}\cdot {\bf{\hat{v}}}^{\perp}\right)\left({\bf{\hat{S}}}_{N}\cdot \frac{{\bf{\hat{q}}}}{m_N}\right)$ \\
        $\hat{\mathcal{O}}_{14} = i\left({\bf{\hat{S}}}_{\chi}\cdot \frac{{\bf{\hat{q}}}}{m_N}\right)\left({\bf{\hat{S}}}_{N}\cdot {\bf{\hat{v}}}^{\perp}\right)$ \\
        $\hat{\mathcal{O}}_{15} = -\left({\bf{\hat{S}}}_{\chi}\cdot \frac{{\bf{\hat{q}}}}{m_N}\right)\left[ \left({\bf{\hat{S}}}_{N}\times {\bf{\hat{v}}}^{\perp} \right) \cdot \frac{{\bf{\hat{q}}}}{m_N}\right] $  \\
        $\hat{\mathcal{O}}_{17}=i \frac{{\bf{\hat{q}}}}{m_N} \cdot \mathbf{\mathcal{S}} \cdot {\bf{\hat{v}}}^{\perp} \mathds{1}_N$ \\
$\hat{\mathcal{O}}_{18}=i \frac{{\bf{\hat{q}}}}{m_N} \cdot \mathbf{\mathcal{S}}  \cdot {\bf{\hat{S}}}_{N}$ \\
    \hline
    \end{tabular*}
    \caption{Quantum mechanical operators defining the non-relativistic effective theory of dark matter-nucleon interactions~\cite{Fitzpatrick:2012ix}.~Here we adopt the notation introduced in Sec.~\ref{eftdm}.~Standard spin-independent and spin-dependent interactions correspond to the operators $\hat{\mathcal{O}}_{1}$ and $\hat{\mathcal{O}}_{4}$, respectively. The operators $\hat{\mathcal{O}}_{17}$ and $\hat{\mathcal{O}}_{18}$ can only arise for spin 1 dark matter, and $\mathbf{\mathcal{S}}$ is a symmetric combination of spin 1 polarisation vectors~\cite{Dent:2015zpa}.~Following~\cite{Fitzpatrick:2012ix}, here we do not consider the operators $\hat{\mathcal{O}}^i_{2}$ and $\hat{\mathcal{O}}_{16}^i$ (see text above Eq.~(\ref{eq:rate}) for further details).~For simplicity, we omit the nucleon index in the operator definitions.}
    \label{tab:operators}
\end{table}

In a dark matter direct detection experiment, the differential rate of nuclear recoil events per unit detector mass is given by:
\begin{align}
\frac{{\rm d} R}{{\rm d} E_R} = \sum_T \xi_T \frac{\rho_\chi}{m_\chi m_T} \int_{|\vec v| \ge v_{\rm min}} {\rm d}^3 v \, |\vec v| f(\vec v) \, \frac{{\rm d}\sigma_T}{{\rm d} E_R} (v^2, E_R) \,,
\label{eq:rate}
\end{align}
where $v_{\rm min}=\sqrt{2 m_T E_R}/(2 \mu_T)$ is the minimum dark matter velocity required to deposit an energy $E_R$ in the detector, $\mu_T$ and $m_T$ are the dark matter-nucleus reduced mass and target nucleus mass, respectively, and $m_\chi$ is the dark matter mass.~In Eq.~(\ref{eq:rate}), $\rho_\chi$ is the local dark matter density, while $f(\vec v)$ is the dark matter velocity distribution in the detector rest frame.~The sum in Eq.~(\ref{eq:rate}) runs over all elements in the detector. Each contribution is weighted by the corresponding mass fraction $\xi_T$.

In Eq.~(\ref{eq:rate}), the differential  cross section for dark matter-nucleus scattering, ${\rm d}\sigma_T/{\rm d} E_R$, depends on the isoscalar and isovector coupling constants, $c_j^0$ and $c_j^1$, respectively, and on nuclear matrix elements of $\mathscr{H}_{\chi T}$.~For an explicit expression, see~\cite{Anand:2013yka}. This very general description of the dark matter-nucleus scattering captures most of the particle physics scenarios that one can conceive.~Important exceptions include models where the dark matter-nucleus scattering is inelastic~\cite{Barello:2014uda}, or scenarios where dark matter-nucleon interactions are mediated by particles with mass comparable or lighter than typical momentum transfers~\cite{Kahlhoefer:2017ddj}. In addition, Eq.~(\ref{eq:H}) cannot be used to describe effects related to meson exchange in nuclei, e.g.~the ``pion pole''~\cite{Hoferichter:2015ipa,Bishara:2016hek}.~However, such effects are known to be important only for momentum transfers comparable with the pion mass, and are therefore expected to be negligible in the mass range of interest for the present analysis~\cite{Bishara:2017nnn}. Finally, Eq.~(\ref{eq:H}) cannot account for operator mixing effects induced by the running of coupling constants. These can be predicted within ultraviolet complete models~\cite{Crivellin:2014qxa}.

Let us now comment on some of the assumptions made while evaluating Eq.~(\ref{eq:rate}).~Regarding the local dark matter density, we adopt the standard value of \unit[0.3]{\GeV/cm$^3$}, although slightly larger values are favoured by astronomical data, e.g.,~\cite{Catena:2009mf}.~For the dark matter velocity distribution in the detector rest frame, we assume a Maxwellian velocity distribution with a circular speed of \unit[220]{km/s} for the local standard of rest and a galactic escape velocity of \unit[544]{km/s} (i.e.~the so-called Standard Halo Model~\cite{Freese:2012xd}).~As far as the detector composition is concerned, here we consider the contribution of Oxygen and Calcium to the scattering cross section, but neglect Tungsten.~For Tungsten, the nuclear response functions, or ``form factors'', associated with (most of) the operators in Tab.~\ref{tab:operators} are currently not known.~Since the predominant isotopes of Oxygen and Calcium have spin $0$, and we neglect Tungsten, only the operators $\mathcal{O}_1$, $\mathcal{O}_3$ $\mathcal{O}_5$, $\mathcal{O}_8$, $\mathcal{O}_{11}$, $\mathcal{O}_{12}$ and $\mathcal{O}_{15}$ contribute to the event rate in the present analysis.~In the notation of~\cite{Fitzpatrick:2012ix}, these operators generate the nuclear responses $W_M^{\tau,\tau'}$, $W_{\Phi''}^{\tau,\tau'}$ and $W_{M\Phi''}^{\tau,\tau'}$. In the zero momentum transfer limit, $W_M^{\tau,\tau'}$ measures the nucleon content of the nucleus (and is proportional to the standard spin-independent form factor), whereas $W_{\Phi''}^{\tau,\tau'}$ measures the nucleon spin orbit coupling content of the nucleus. Finally, the nuclear response $W_{M\Phi''}^{\tau,\tau'}$ arises from the interference of the nuclear currents underlying $W_M^{\tau,\tau'}$ and $W_{\Phi''}^{\tau,\tau'}$.

Conclusions based on Eq.~(\ref{eq:rate}) are affected by uncertainties in astrophysical and nuclear physics inputs.~For kinematical reasons, only uncertainties on the nuclear response functions at zero momentum transfer are relevant for light dark matter.~Whereas $W_{M}^{\tau\tau'}(0)$ is known exactly, being proportional to the square of the number of nucleons in the nucleus, uncertainties on $W_{\Phi''}^{\tau,\tau'}(0)$ and $W_{M\Phi''}^{\tau,\tau'}(0)$ must be assessed through nuclear structure calculations.~In the case of Helium, the relative uncertainty on these response functions was found to be of a factor of 3 or so using an ab initio no core shell model approach~\cite{Gazda:2016mrp}.~Using large-scale nuclear structure calculations, similar results were found for the nuclear response functions of interest in the case of Xenon isotopes~\cite{Klos:2013rwa}.~On the other hand, astrophysical uncertainties can play a crucial role in the search for light dark matter, especially those on the dark matter velocity distribution.~For example, in the small mass limit, it has been found that exclusion limits can be modified by up to few orders of magnitude by variations in the astrophysical inputs that govern the dark matter and baryon mass profiles in the Milky Way~\cite{Benito:2016kyp}.~In order to consistently compare our results with those in~\cite{Schneck:2015eqa}, we will present our exclusion limits focusing on the Standard Halo Model, and adopting the nuclear response functions for Oxygen and Calcium computed in~\cite{Catena:2015uha} through a shell model calculation.

\section{The CRESST Experiment}\label{cresst}

CRESST (Cryogenic Rare Event Search with Superconducting Thermometers) is a direct dark matter search experiment. The anticipated dark matter signals are nuclear recoils in a scintillating calcium tungstate (CaWO$_4$) target crystal. The target detectors are operated at a temperature of around \unit[15]{mK}. To shield the experiment from background signals, mainly induced by cosmic rays, the experiment is located at the underground laboratory of the LNGS (Laboratori Nazionali del Gran Sasso) in central Italy. This work uses data from phase 2 of CRESST-II\footnote{Recently an upgrade of the experiment has been performed. CRESST-III started taking data in July 2016 (until February 2018).}, which started in July 2013 and ended in August 2015. In total 18 detector modules with an overall mass of $\sim$ \unit[5]{kg} were operated. \cite{commissioning,results}

The module \textit{Lise} was the detector module with the lowest trigger threshold for nuclear recoils (\unit[0.307]{keV}). Limits on the elastic spin-independent dark-matter-nucleon cross-section from this detector module were published in 2016 (see \cite{results}). For dark matter masses below \unit[2]{\GeV} these limits led the field at the time of the publication.\footnote{They were surpassed in parts of this mass region by first results from CRESST-III \cite{Petricca:2017zdp}.} This work uses the same data as \cite{results}.

The detector module consists of a scintillating CaWO$_4$ crystal with a mass of \unit[300]{g} (phonon detector) and an independent light detector. Most of the energy deposited by recoils leads to a phonon signal, which is thus used for the energy determination. The fraction of the energy that yields a scintillation light signal is called light yield and is used for discrimination between different type of recoils: The scattered particle can either be a calcium, tungsten or oxygen nucleus, or an electron. While most of the background events are electron recoils, the anticipated dark matter signals are nuclear recoils.

\begin{figure}[h]
	\centering
  \includegraphics[width=0.45\textwidth]{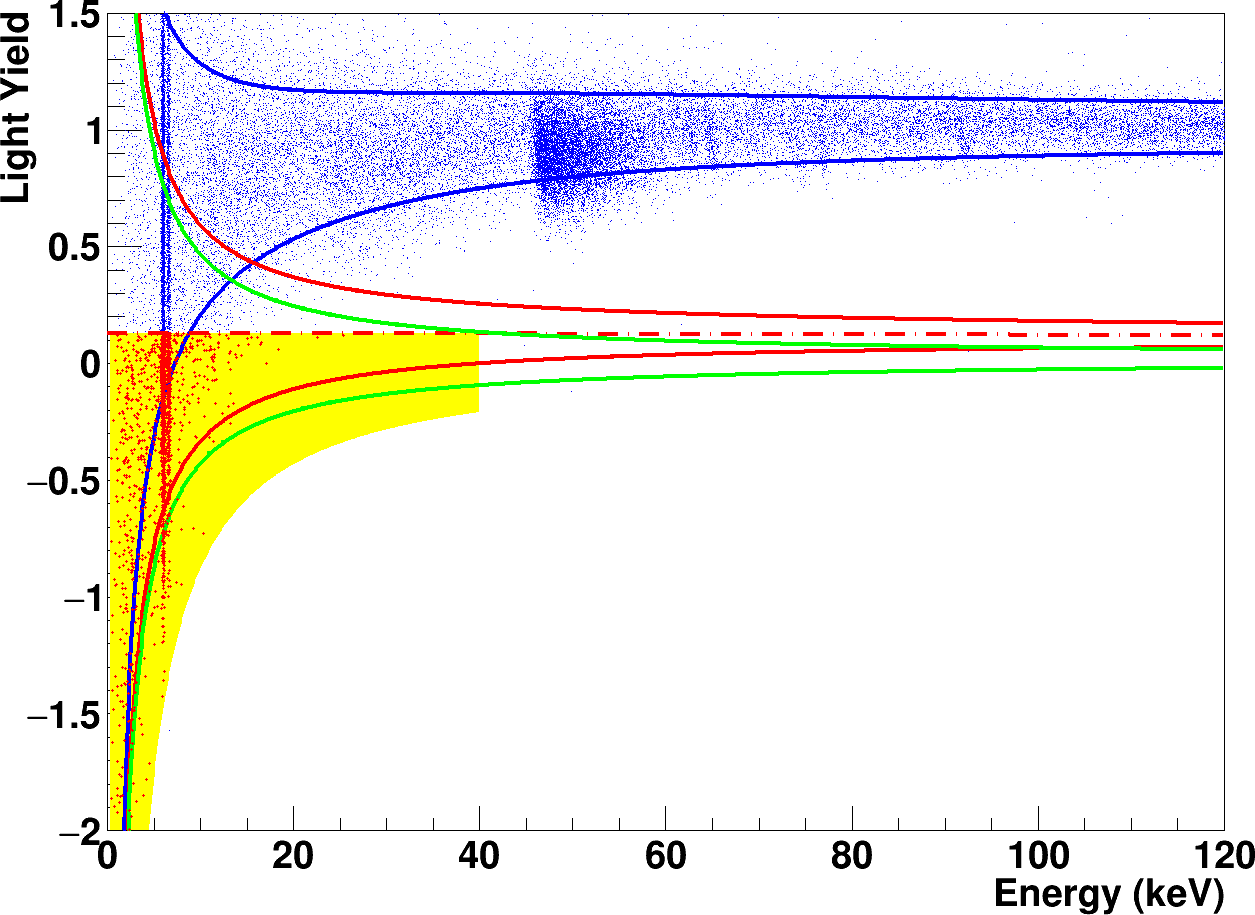}
	\caption{All events from the detector module \textit{Lise} after all cuts. The light yield, the fraction of light to phonon signal, is plotted against the (phonon) energy. Solid lines are 90 \% boundaries for electron recoils (blue) and nuclear recoils off oxygen (red) and tungsten (green). The red dashed line is the center of the oxygen band. The yellow area is the acceptance region.}
	\label{lightyield}
\end{figure}

\begin{figure}[h]
	\centering
  \includegraphics[width=0.45\textwidth]{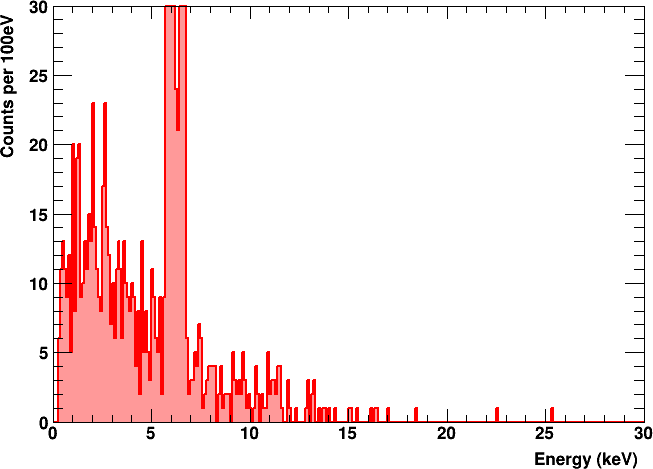}
	\caption{Energy spectrum of all events in the previously defined acceptance region from the detector module \textit{Lise}. The range in the y-axis is chosen for reasons of clarity, although a few bins surpass the upper bound.}
	\label{allevents}
\end{figure}

Figure \ref{lightyield} shows the dataset used for this work. For more details on data preparation and cuts see \cite{results}. The light yield is normalized to 1 for electron recoils with an energy of \unit[122]{keV}. Since only nuclear recoils are considered as a possible dark matter signature in this analysis, the acceptance region is chosen accordingly: To avoid leakage from electron recoils, the upper bound is defined as the center of the oxygen band, which is the highest of the nuclear recoil bands in figure \ref{lightyield}. The lower bound is chosen to be the 99.5 \% lower boundary of the lowest band, which is the tungsten band. The calcium band (not shown in fig. \ref{lightyield}) lies between the oxygen and the tungsten band. In terms of energy the acceptance region spans from threshold (\unit[307]{eV}) to \unit[40]{keV}. The acceptance region, as well as all methods of data preparation and selection, have been defined and fixed before unblinding the data. Figure \ref{allevents} shows the energy distribution of all events in the acceptance region. The double peak at \unit[6.0]{keV} and \unit[6.6]{keV} is due to a $^{55}$Fe X-ray source that was installed for the calibration of a detector module close to \textit{Lise}. Although this was an unintentional exposure, it doesn't significantly influence the sensitivity of the experiment because of the narrow width of the peaks. The smaller peaks at \unit[2.7]{keV} and \unit[8.1]{keV} are also understood. They originate from cosmogenic activation of tungsten and copper fluorescence respectively.

\section{Effective Field Theory Data Analysis}\label{analysis}

The goal of this analysis is to set limits on the coupling constants $c_i$ defined in equation \ref{eq:H}. In principle the theory allows any linear combination of the operators, but we restrict ourselves to limits on single operators individually. The calculation of the expected spectra for each operator and for each isotope was executed based on a Matlab code released by the CDMS collaboration for a similar analysis \cite{Schneck:2015eqa}. For each operator, the spectra are then added up according to the abundance of each isotope in calcium tungstate (CaWO$_4$).

Only the most abundant isotopes of oxygen ($^{16}$O) and Calcium ($^{40}$Ca) are taken into account. The nuclear form factors from section \ref{eftdm} have already been calculated for these two isotopes using shell model computations \cite{Catena:2015uha}, but not for the much heavier tungsten isotopes. Therefore, tungsten is left out of this analysis.

In order to calculate limits, Yellin's optimum interval method is used \cite{yellin1,yellin2}. This method doesn't require a background model and the implementation of different spectrum shapes is unproblematic. Also, without taking into account all isotopes in the spectrum calculation (which leads to lower expected spectra), limit calculation is still possible and valid, but yields conservative results.

\section{Results and Conclusions}\label{conclusions}

\begin{figure*}[h]
\centering
\includegraphics[width=0.35\textwidth]{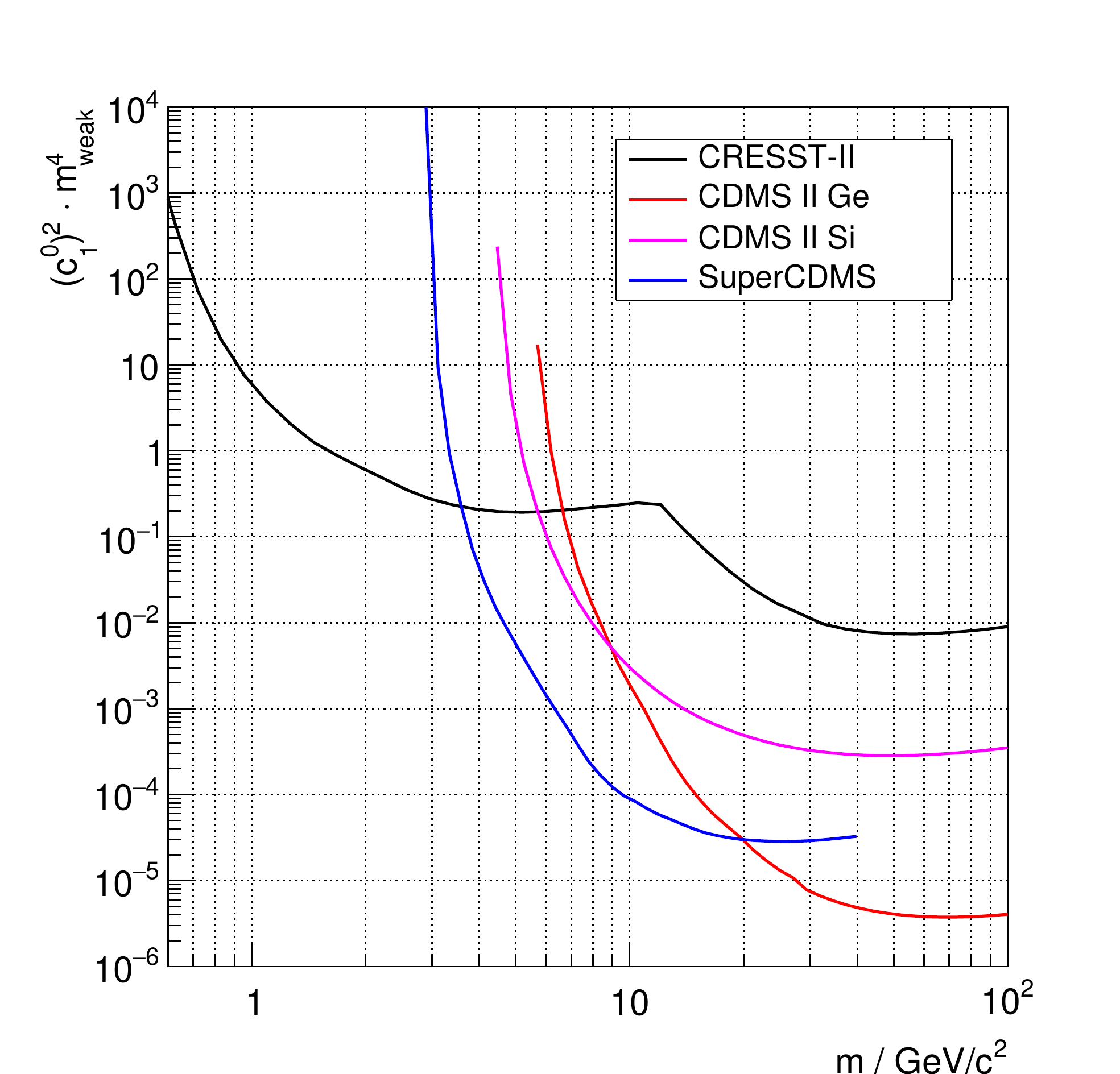}
\includegraphics[width=0.35\textwidth]{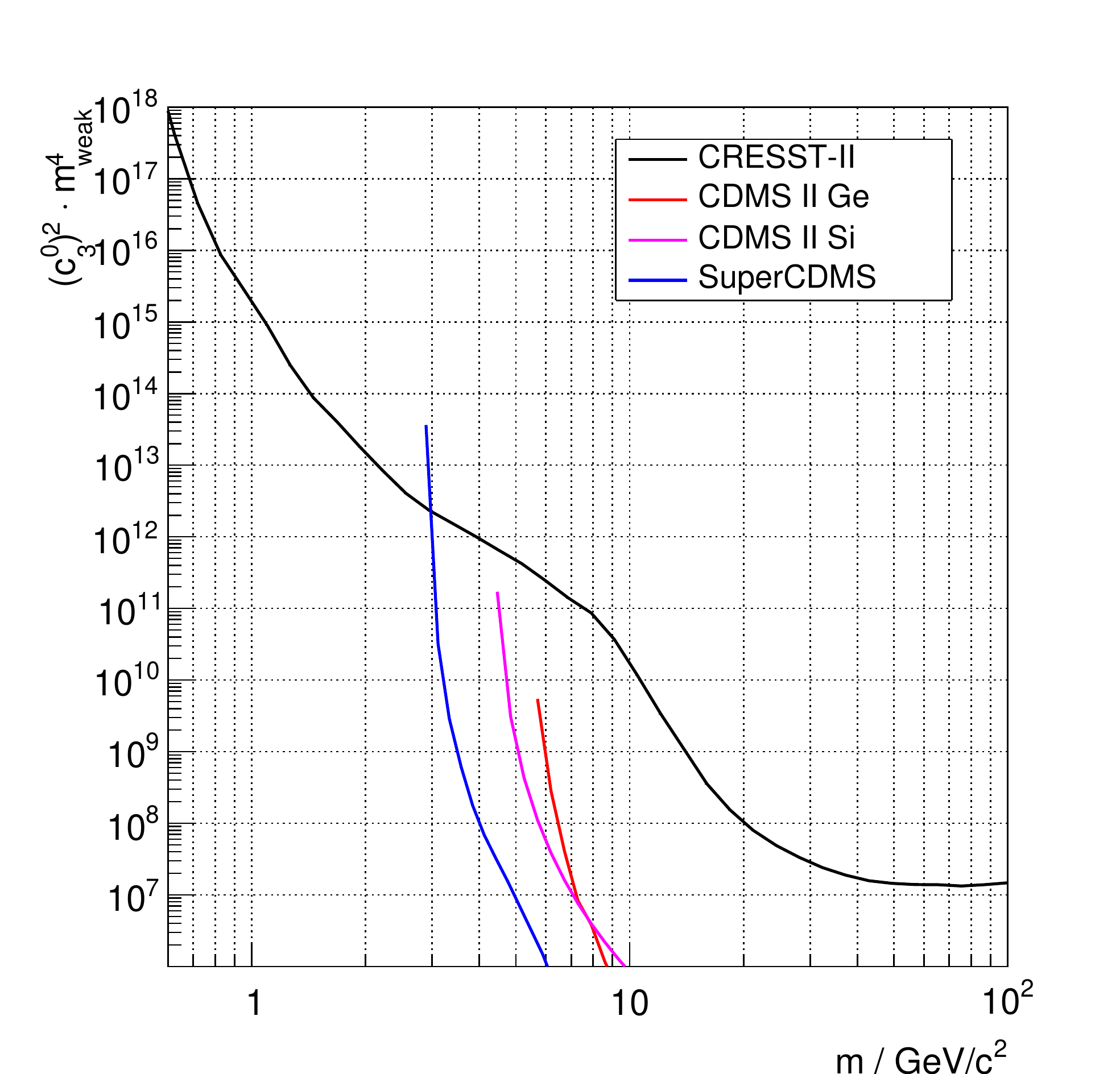}
\includegraphics[width=0.35\textwidth]{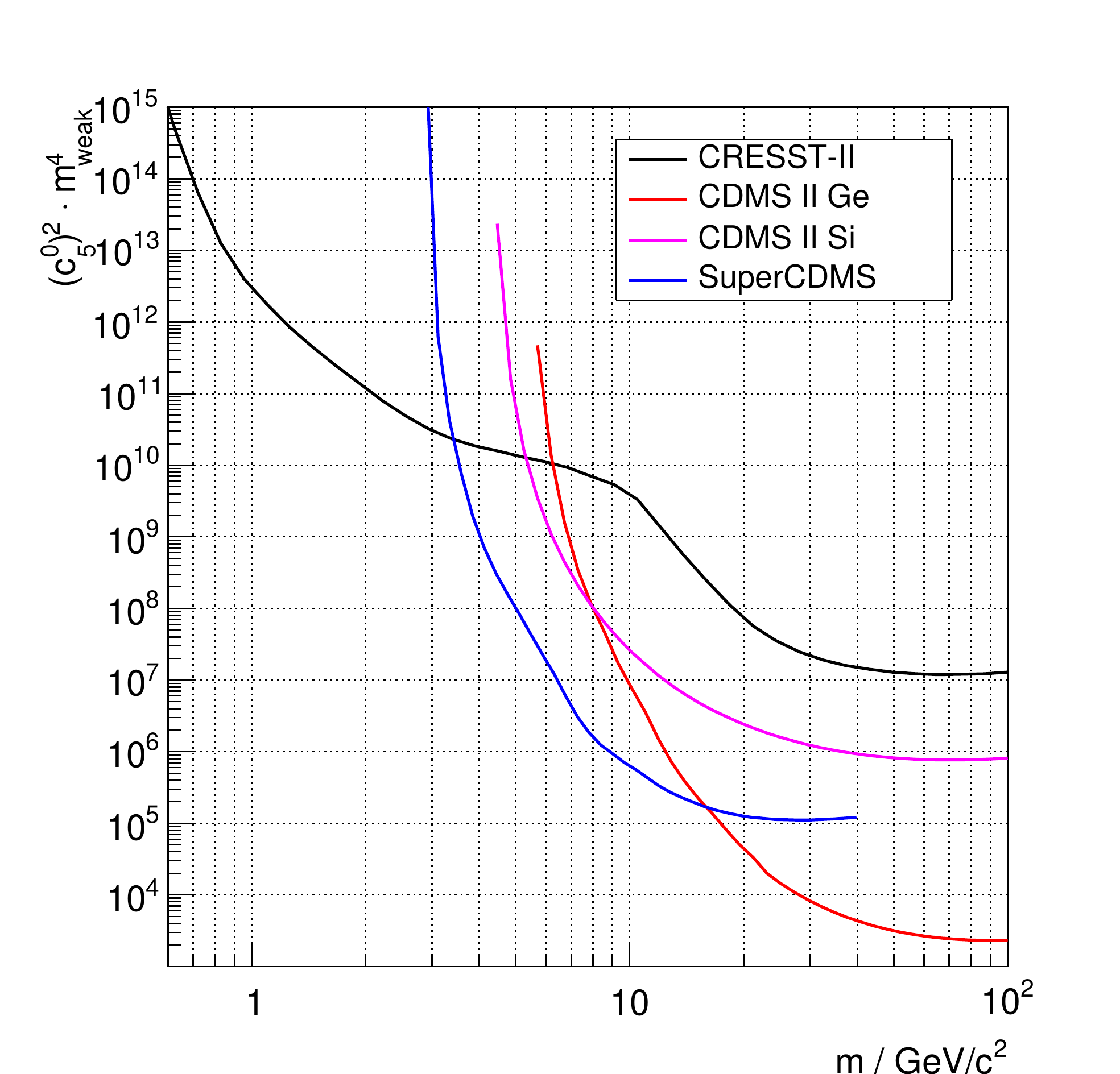}
\includegraphics[width=0.35\textwidth]{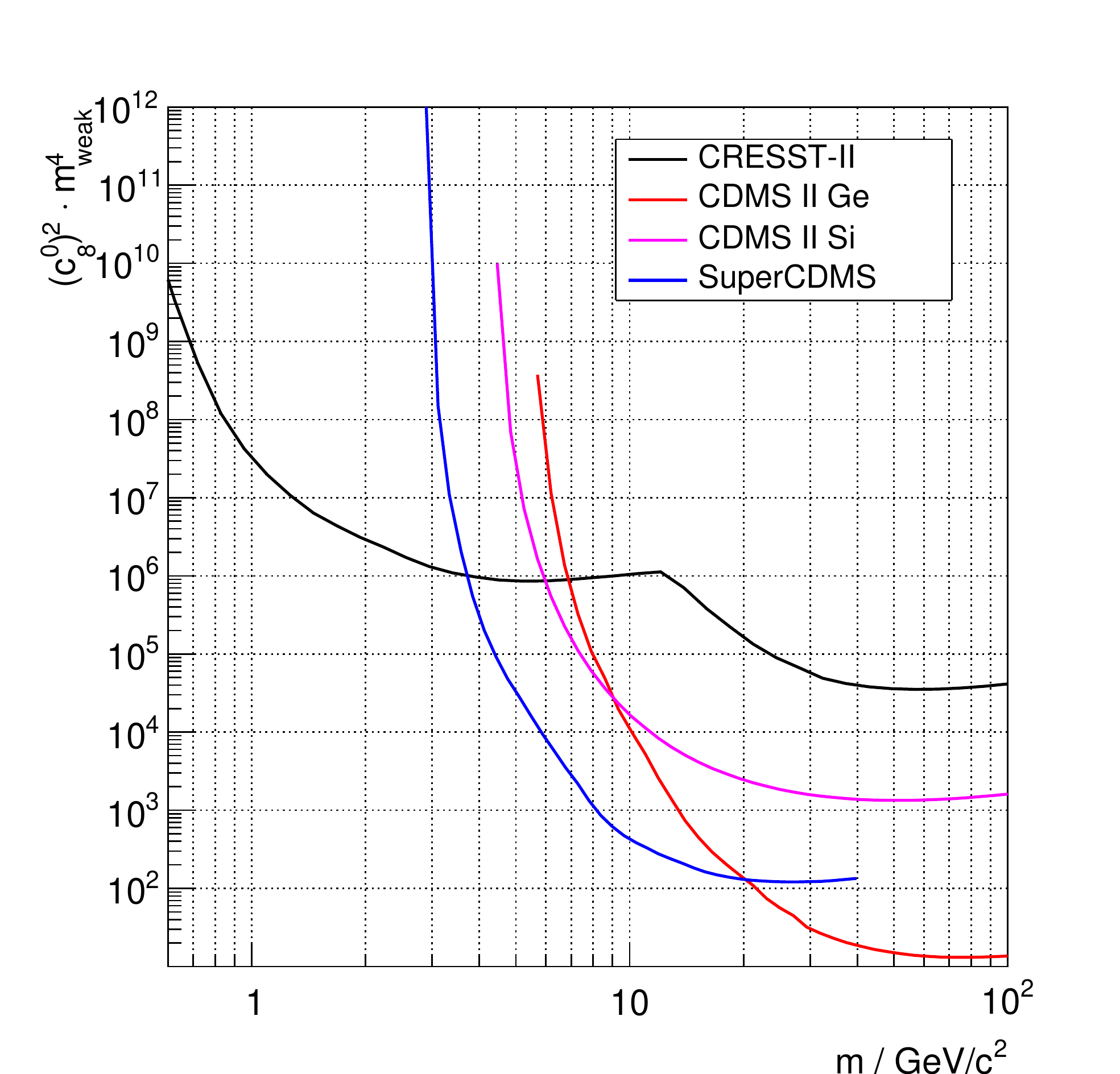}
\includegraphics[width=0.35\textwidth]{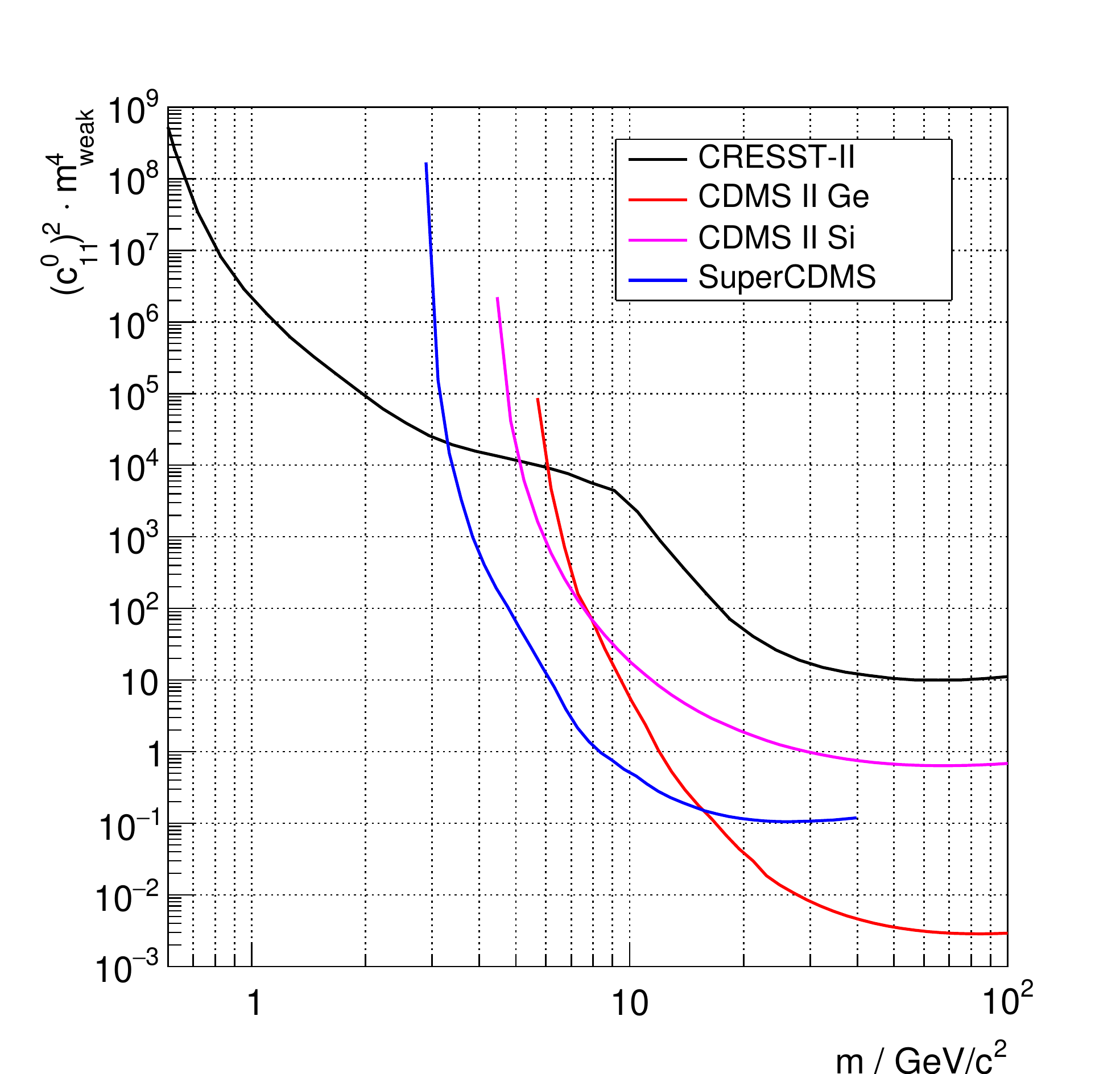}
\includegraphics[width=0.35\textwidth]{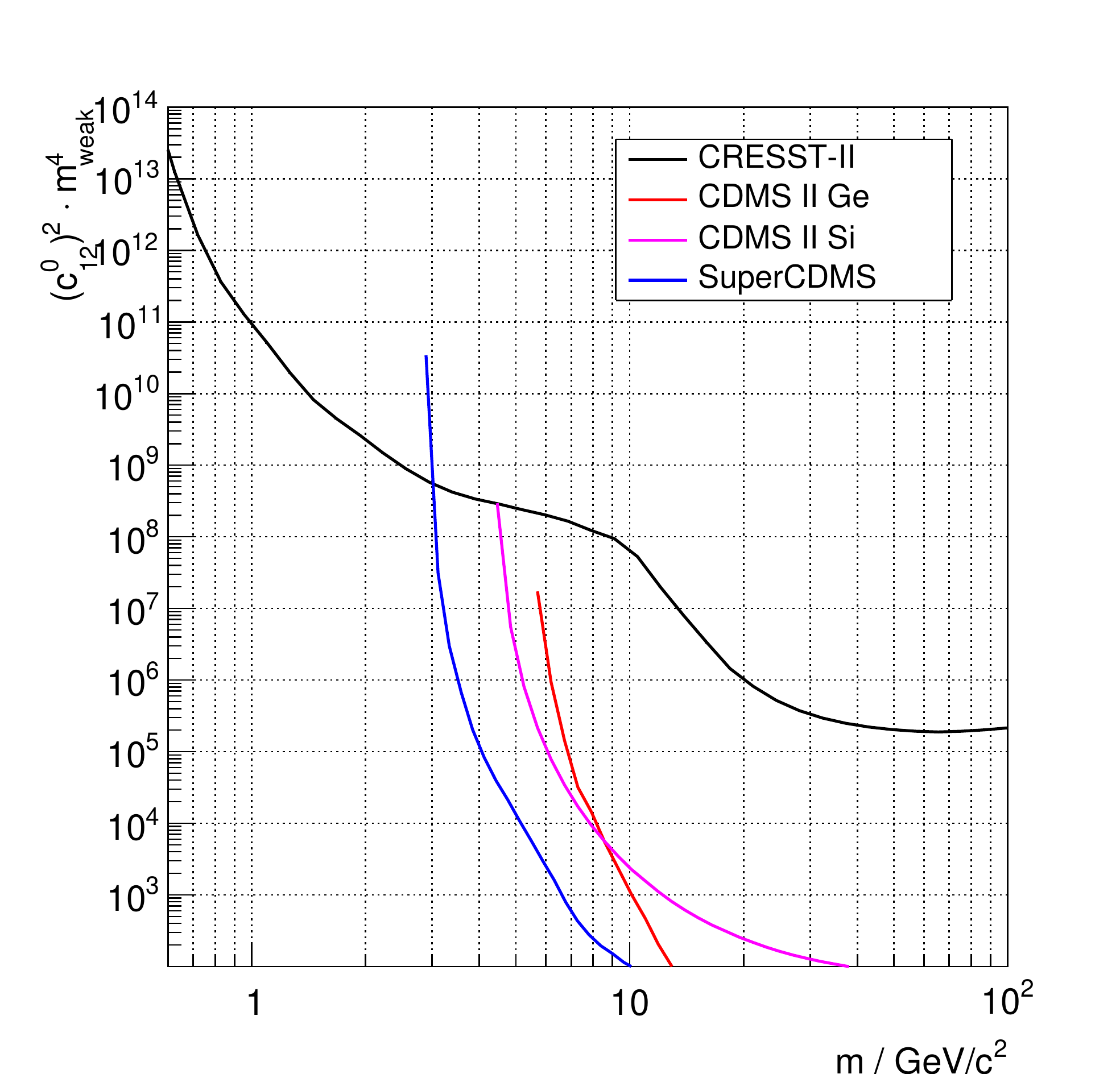}
\includegraphics[width=0.35\textwidth]{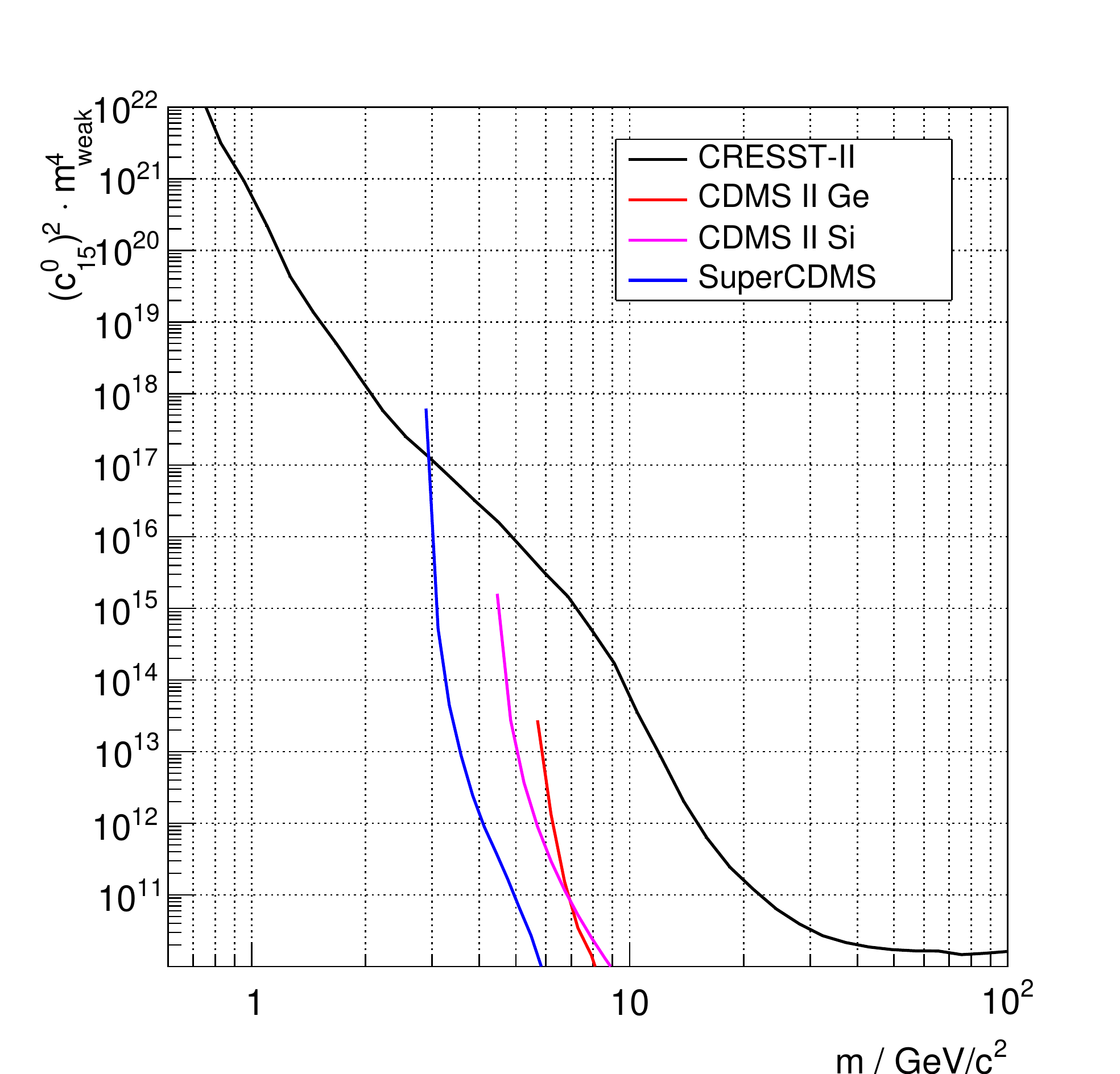}
\caption{Upper 90\% limits for the Wilson coefficients $c_1^0$, $c_3^0$, $c_5^0$, $c_8^0$, $c_{11}^0$, $c_{12}^0$ and $c_{15}^0$ as a function of the dark matter particle mass (black), compared to results from CDMS \cite{Schneck:2015eqa} (other colors).}\label{results}
\end{figure*}

The limits for the Wilson coefficients $c_1^0$, $c_3^0$, $c_5^0$, $c_8^0$, $c_{11}^0$, $c_{12}^0$ and $c_{15}^0$ are shown in fig. \ref{results}. The results are compared to limits from the CDMS experiment \cite{Schneck:2015eqa}. An effective field theory analysis for the Xenon100 experiment has also been published \cite{Aprile:2017aas}, but only covers dark matter masses above \unit[10]{\GeV} and is therefore not relevant for comparison here.

The isotopes that are taken into account for this analysis, $^{16}$O and $^{40}$Ca, both contain the same number of protons and neutrons. Consequently, we are not sensitive to isovector operators in this analysis. Also, the overall spin of both of the two isotopes is zero. For this reason, the strongest limits, compared to the CDMS experiment, are obtained for the coefficients $c_i^0$ for operators that contain no dependence on the nucleon spin, namely the operators $\mathcal{O}_1$, $\mathcal{O}_5$, $\mathcal{O}_8$ and $\mathcal{O}_{11}$. However, limits on the coeficcients $c_i^0$ for the operators $\mathcal{O}_{3}$, $\mathcal{O}_{12}$ and $\mathcal{O}_{15}$ are also provided.

Fig. \ref{results} shows that this present analysis sets leading limits on all of the above-mentioned Wilson coefficients for dark matter masses below \unit[3-4]{\GeV}. Similar to the standard spin-independent analysis \cite{results}, the low energy threshold of the CRESST detectors leads to particularly high sensitivity for low particle masses. The upgrade to CRESST-III, with a threshold that is even lower and improvements on the detector design, holds considerable potential to further improve these limits, especially for low dark matter particle masses. It should also be noted that in the effective field theory, where differences in nuclear properties of the target nuclei play a decisive role, the comparison of multiple different complementary experiments becomes even more important than in the standard spin-independent analysis.


\bibliographystyle{h-physrev}

\end{document}